\documentclass[conference]{IEEEtran}
\IEEEoverridecommandlockouts
\usepackage{cite}
\usepackage{amsmath,amssymb,amsfonts}
\usepackage{algorithmic}
\usepackage{graphicx}
\usepackage{textcomp}
\usepackage{xcolor}
\usepackage{booktabs}
\usepackage{caption}
\usepackage{xcolor}
\usepackage{colortbl}
\usepackage{makecell}
\usepackage{array}
\usepackage{dblfloatfix}    % To enable figures at the bottom of page
\usepackage{hyperref}

\definecolor{orange}{rgb}{1,0.5,0}
\definecolor{lightred}{rgb}{1,0.8,0.8}
\definecolor{green}{rgb}{0.8,1,0.8}
\definecolor{purple}{rgb}{0.8,0.6,0.8}

\def\BibTeX{{\rm B\kern-.05em{\sc i\kern-.025em b}\kern-.08em
    T\kern-.1667em\lower.7ex\hbox{E}\kern-.125emX}}
\begin{document}

% \title{Optical Complex-Valued Neural Networks via Photoelectric Multiplication
% \title{Complex-Valued Photoelectric Neural Networks for Time and Energy Advantages
% \title{Doubly Fast, Quadratically Efficient\\ Optical QAM Neural Networks
\title{QAMNet: Fast and Efficient\\ Optical QAM Neural Networks
% \thanks{\textcolor{red}{TODO. M.G.B. was supported by a grant from NTT Research.}}
}

\author{\IEEEauthorblockN{Marc Gong Bacvanski}
\IEEEauthorblockA{\textit{MIT RLE} \\
marcbac@mit.edu}
\and
\IEEEauthorblockN{Sri Krishna Vadlamani}
\IEEEauthorblockA{\textit{MIT RLE}}
\and
\IEEEauthorblockN{Kfir Sulimany}
\IEEEauthorblockA{\textit{MIT RLE}}
\and
\IEEEauthorblockN{Dirk Robert Englund}
\IEEEauthorblockA{\textit{MIT RLE}
}}

\maketitle

\begin{abstract}
The energy consumption of neural network inference has become a topic of paramount importance with the growing success and adoption of deep neural networks (DNNs). 
Analog optical neural networks (ONNs) can reduce the energy of matrix-vector multiplication in neural network inference below that of digital electronics. However, realizing this promise remains challenging due to digital-to-analog (DAC) conversion — even at low bit precisions $b$, encoding the $2^b$ levels of digital weights and inputs into the analog domain requires specialized and power-hungry electronics. 
Faced with similar challenges, the field of telecommunications has developed the complex-valued Quadrature-Amplitude Modulation (QAM), the workhorse modulation format for decades. 
QAM maximally exploits the complex amplitude to provide a quadratic \mbox{$O(N^2)\to O(N)$} energy saving over intensity-only modulation.
Inspired by this advantage, this work introduces QAMNet, an optical neural network hardware and architecture with superior energy consumption to existing ONNs, that fully utilizes the complex nature of the amplitude of light with QAM.
When implemented with conventional telecommunications equipment, we show that QAMNet accelerates complex-valued deep neural networks with accuracies indistinguishable from digital hardware, based on physics-based simulations.
Compared to standard ONNs, we find that QAMNet ONNs: (1) attain higher accuracy above moderate levels of total bit precision, (2) are more accurate above low energy budgets, and (3) are an optimal choice when hardware bit precision is limited.
\end{abstract}


\section{Introduction}\label{sec:intro}

\begin{figure*}
    \centering
    \includegraphics[width=1\linewidth]{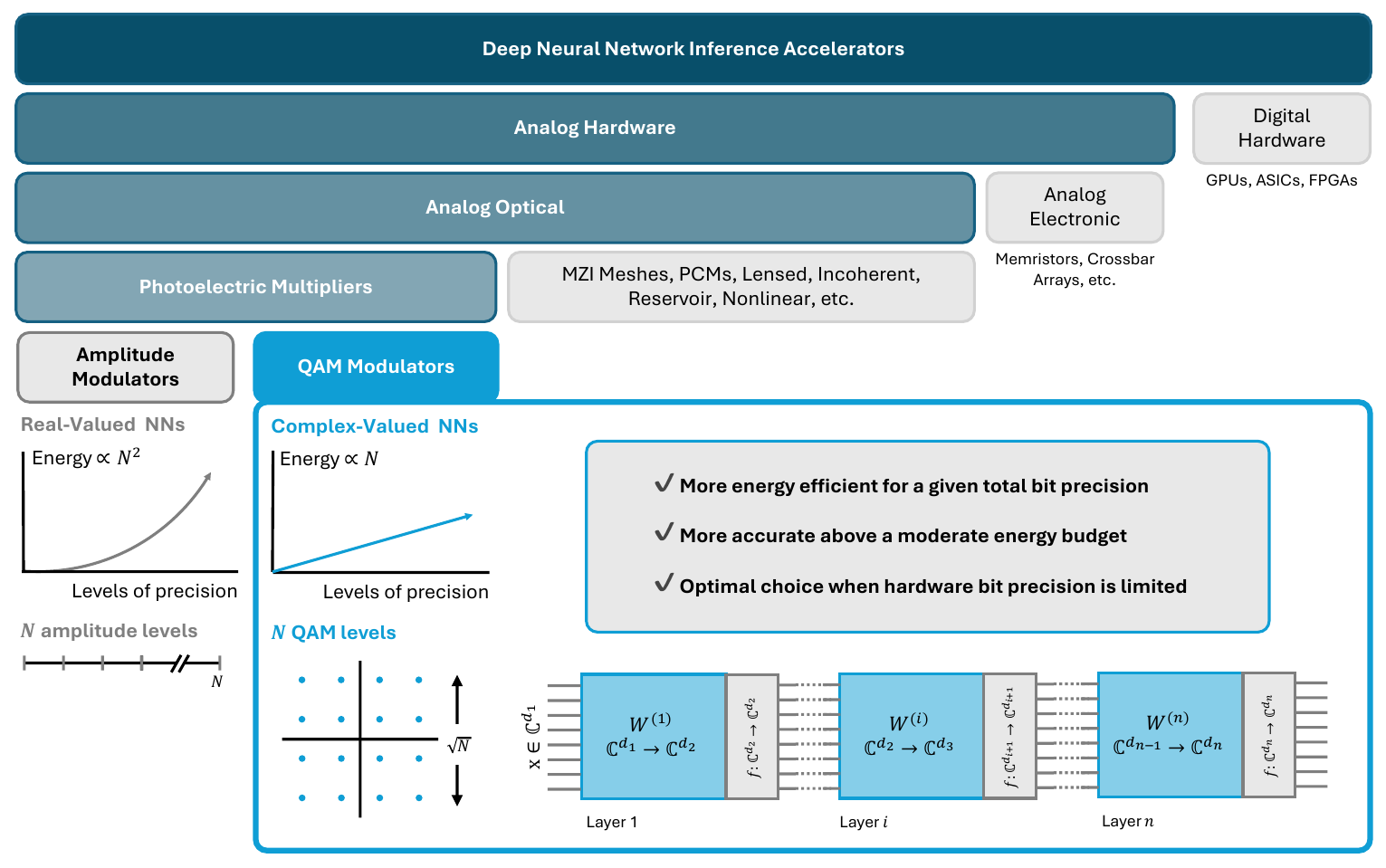}
    \caption{Taxonomy of DNN inference hardware, highlighting the advantages of Quadrature-Amplitude Modulation (QAM) optical neural networks and their relationship to other analog and optical deep neural network accelerator approaches. QAMNet is a QAM-based photoelectric multiplication scheme that provides linear scaling of energy with respect to levels of precision.}
    \label{fig:intro_figure}
\end{figure*}

\begin{figure*}
    \centering
    \includegraphics[width=1\linewidth]{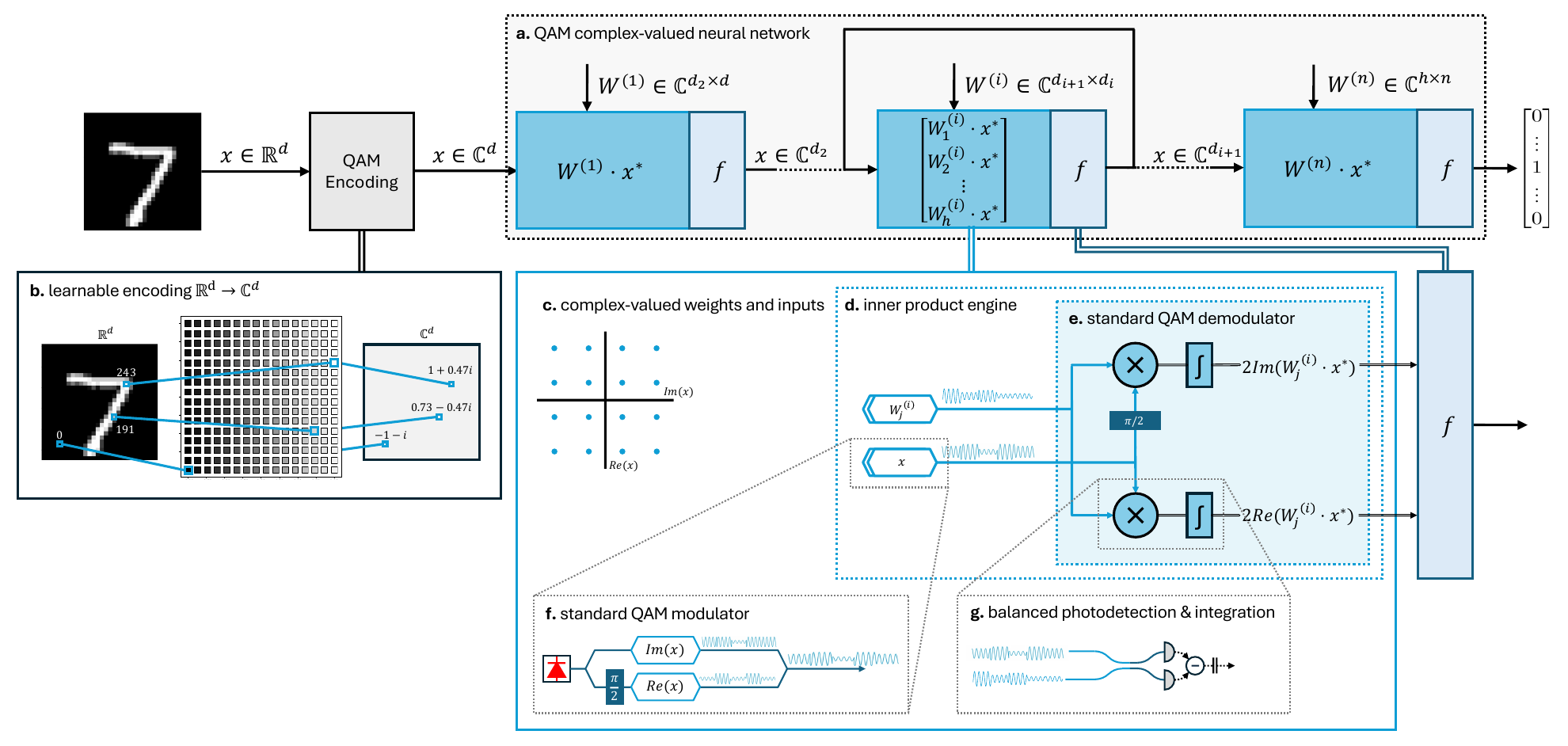}
    \caption{Illustration of QAMNet's algebra, circuit, and hardware. \textbf{a.} Algebra of complex-valued multi-layer perceptron neural network inference. A linear layer $i$ with input dimension $d_i$ and output dimension $d_{i+1}$ is parameterized by weight matrices $W^{(i)}\in\mathbb C^{d_{i+1}\times d_{i}}$ and a nonlinear function $f:\mathbb C^{d_{i+1}}\to\mathbb C^{d_{i+1}}$. Layer $i$ shows detail of the inner products computed during inference. \textbf{b.} Detail of the learnable encoding that maps real-valued data to complex-valued data of the same dimension. \textbf{c.} Detail of the computation of a single layer. Complex-valued weights and inputs represented as symbols in a QAM constellation. \textbf{d.} Detail of the I/Q photoelectric multiplier computing a complex-valued inner product. \textbf{e.} Components of the I/Q photoelectric multiplier that are found in standard QAM demodulators, with the only difference being the replacement of the local oscillator with a I/Q modulator for weights. \textbf{f.} Detail of each I/Q QAM modulator, highlighting how two amplitude modulators (i.e. implemented as Mach-Zehnder modulators) are used off-phase to perform phase and amplitude modulation. \textbf{g.} Detail of the mixer found in standard QAM demodulators, implemented by a beamsplitter (depicted as an evanescent coupler) and balanced photodetectors which collect the difference current on a capacitor.}
    \label{fig:main_figure}
\end{figure*}

Despite the recent successes and rapid adoption of deep neural networks (DNNs)~\cite{lecun2015deep, vinyals2019grandmaster, brown2020language}, their use is limited by the high energy consumption required for inference. The power consumption of DNN inference is in large part due to the repeated multiply-accumulate (MAC) operations required to perform matrix-vector multiplication~\cite{sze2017efficient, yang2017method, patterson2021carbon}. Therefore, both specialized digital hardware including ASICs~\cite{chen2016eyeriss, jouppi2017datacenter}, FPGAs~\cite{sharma2016high}, and GPUs, as well as analog hardware have been developed to perform these operations faster and with lower energy. We show a taxonomy of accelerator architectures in Figure \ref{fig:intro_figure}, with emphasis on the design space decisions that motivate the exploration of analog optical accelerators.

Analog computing offers a compelling alternative to digital hardware due to opportunities for inherent parallelism~\cite{hamerly2021edge}, lower energy consumption not limited by transistor switching energy~\cite{wang2022optical}, and integration with co-designed sensors~\cite{lin2018all}. 
Analog \textit{electronic} methods include memristor crossbar arrays~\cite{hu2018memristor, xia2019memristive, wang2017memristors}, phase change memory~\cite{burr2015experimental, zhang2019designing} and capacitor-based schemes~\cite{xia2016technological}. On the other hand, analog \textit{optical} computing offers new paradigms of parallelism,
% and is aided by the miniaturization and scalability gains of integrated photonics. These methods 
and includes approaches like Mach-Zehnder Inteferometer (MZI) meshes, lensed free-space schemes, nonlinear optical materials, and particularly, photoelectric multipliers~\cite{shen2017deep, wetzstein2020inference, shastri2021photonics, bandyopadhyay2022single, feldmann2021parallel, hamerly2019large, sludds2022delocalized, sulimany2024quantum}, which we will focus on in this paper.

Analog implementations of neural networks require quantization, where digitally-represented ``full-precision" values must be discretized into a fixed and typically lower number of levels that the hardware can implement~\cite{gupta2015deep}. Each matrix and vector element value is usually represented as a signal modulated in intensity or real-valued amplitude (positive and negative). Values represented in analog, despite being continuous in theory, must respect a minimum distinguishable step size between two signal levels. This step size $\Delta$ is lower-bounded by the DAC's noise level, since if two signal levels differ by less than $\Delta$ units of energy, they will be indistinguishable in the presence of noise of amplitude $\Delta$. Therefore, in real amplitude modulation schemes, representing $N$ unique levels requires $\left(\frac{N-1}{2}\Delta\right)^2$ units of energy~\cite{walden1999analog}.

Faced with this quadratic scaling of energy, the field of telecommunications has developed Quadrature-Amplitude Modulation (QAM), which uses quadratically less energy for the same number of symbols. To represent $N$ unique symbols with QAM in the presence of noise $\Delta$ around each symbol requires only $\sqrt N$ levels per axis (real and complex). Therefore, the total energy needed for both modulators is $2\times\left(\frac{\sqrt N - 1}{2}\Delta\right)^2$, which is linear in $N$, in contrast to real-valued amplitude modulation as depicted in Figure \ref{fig:intro_figure}.

QAM modulation is a discretized form of I/Q modulation, where each value (referred to as a symbol in QAM) is defined by a real (in-phase $I$) and imaginary (quadrature $Q$) component, whose resulting phasor on a carrier wave of frequency $\omega$ can be written as $s(t)=I\cos(\omega t)+Q\sin(\omega t)$. Due to the fixed precision of each modulator, a QAM-modulated signal takes on discrete points on the complex plane, which determines the modulator's \textit{constellation diagram}\cite{papen2019lightwave}. 

In this work, we propose QAMNet, a QAM-based optical neural network that offers energy and accuracy advantages for deep learning tasks. We propose a method for I/Q photoelectric multiplication, a scheme that naturally and optimally implements photoelectric multiplication of complex-valued weights and inputs. We show that our scheme is amenable to implementation with standard telecommunications QAM modulators. Applied to deep learning inference tasks, it achieves accurate inference of pre-trained complex-valued neural networks even in the presence of realistic noise. Finally, we compare the energy and accuracy of real-valued neural networks with real-valued amplitude photoelectric multipliers, versus QAMNet with I/Q photoelectric multipliers. We find that QAMNet ONNs: (1) attain higher accuracy above moderate levels of total bit precision, (2) are more accurate above low energy budgets, and (3) are an optimal choice when hardware bit precision is limited.

\section{I/Q Photoelectric Multiplication} \label{sec:photoelectric_multiplication_scheme}
In this section we outline the I/Q modulation-based photoelectric multiplication scheme. We show how it natively supports complex-valued multiply-accumulate operations for efficient complex-valued inner product computation, as well as how the same hardware can be used to perform real-valued inner products twice as fast as the amplitude modulation photoelectric multiplication scheme.

Figure \ref{fig:main_figure} illustrates our proposed complex-valued inner product engine using I/Q encoded values. A complex-valued weight or input is modulated onto a carrier wave in I/Q space, where the \underline{I}n-phase and \underline{Q}uadrature components represent the real and imaginary parts of the value, respectively. This allows each I/Q-encoded symbol to represent a point in the complex plane. A signal in I/Q space can be represented in the time domain as Equation~\ref{iq_modulation_time_domain}, where the in-phase and quadrature components are modulated in amplitude on two out-of-phase carriers of frequency $\omega$, which are combined as shown in Figure \ref{fig:main_figure}.f. The $j$'th element of a complex-valued vector $x$ would be modulated as:
\begin{equation}\label{iq_modulation_time_domain}
    s_j(t)=\text{Re}(x_j)\cos(\omega t)+\text{Im}(x_j)\sin(\omega t)
\end{equation}

Equation~\ref{iq_modulation_time_domain} can be rewritten in phasor notation as Equation~\ref{eq:iq_modulation_phasor}, highlighting the amplitude and phase of the modulated value.
\begin{equation}\label{eq:iq_modulation_phasor}
s_j(t) = A_je^{i(\omega t +\phi_j)}
\end{equation}
\begin{equation}
\begin{split}
A_j = \sqrt{\text{Re}(x_j)^2+\text{Im}(x_j)^2}
&\qquad \phi_j = \arctan\left(\frac{\text{Im}(x_j)}{\text{Re}(x_j)}\right)\nonumber\\
\end{split}
\end{equation}

The weight values $w$ are encoded in the same way. As shown in Figure \ref{fig:main_figure}.d, each I/Q-encoded input and weight value is fanned out into two beam paths. Each beam path leads to balanced photodetection with an electronic integrator, shown in Figure \ref{fig:main_figure}.g. Both beam paths use the same beamsplitter with a transfer matrix of $\frac{1}{\sqrt{2}}\begin{bmatrix}1&1\\1&-1\end{bmatrix}$, but with a $\frac \pi2$ phase shift applied to one of the inputs at the top path. Balanced photodetection takes the difference of the photocurrents, yielding an accumulated charge proportional to $2\text{Im}(w_j x_j^*)$ and $2\text{Re}(w_j x_j^*)$ at the top and bottom beam paths, respectively.

The photocurrents produced by these individual scalar multiplications are integrated over time with a capacitor to yield inner products at the top and bottom beam paths. The charge on the capacitor will be proportional to \mbox{$\sum_{j=1}^n\text{Im}(w_j x_j^*)$} and $\sum_{j=1}^n\text{Re}(w_j x_j^*)$ on the top and bottom beam paths respectively, thereby yielding the real and imaginary components of the desired complex-valued inner product $w\cdot x^*$ (see Appendix~\ref{appendix:iq_inner_products} for full derivation).

\section{Complex-Valued Neural Network Inference with Telecom QAM Modulators} \label{sec:inference_with_telecom_qam_modulators}
In this section, we demonstrate the feasibility of using existing telecommunications hardware for accelerating complex-valued neural network (CVNN) inference. Both the modulators and the mixers used in the I/Q photoelectric multiplier are found in existing telecommunications equipment as QAM modulators and demodulators, respectively. 
As mentioned previously, these digital QAM modulators transmit discrete symbols with discrete $I$ and $Q$ values, the set of which comprises the modulator's constellation, which is typically arranged as a square grid with equal vertical and horizontal spacing, and measured by the number of points on each axis (QAM side).
This fixed set of constellation points means that every complex-valued weight and input activation in the CVNN must be quantized to a constellation point. 
Furthermore, non-idealities in modulators and photodetector response results in noisy output signals. Here, we demonstrate how our I/Q photoelectric multiplier enables accurate inference of a deep CVNN, under realistic noise and quantization figures.

One successful application of complex-valued neural networks is on the RadioML 2016.10A dataset \cite{o2016radio}, which is classification of a sequence of I/Q samples into 11 modulation format classes. We train the ``DSN (Deep Signal Network)" model described in~\cite{tu2020complex}, and then transfer the weights to the physical implementation using a standard post-training quantization strategy as described in \cite{jacob2018quantization}.

Because the real and imaginary components of the inner product (each output element of the matrix multiplication) are read out by separate ADCs, the real and imaginary components of the result receive independent and identically distributed noise. The total noise from the DAC, modulator, photodetectors, and ADC is modeled as a Gaussian random variable with zero mean and a standard deviation given by~$\sigma_\text{noise}=\sigma_\text{signal}/\sqrt{\text{SNR}}$.
\begin{figure}
    \centering
    \includegraphics[width=1\linewidth]{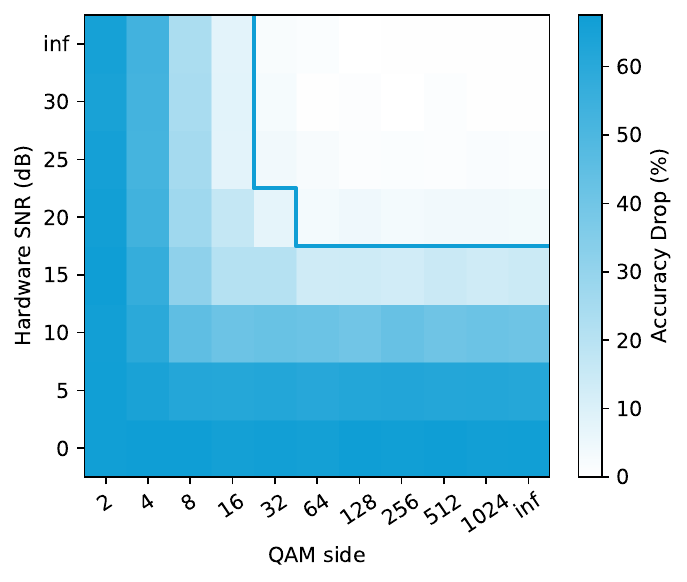}
    \caption{Accuracy degradation across hardware Signal to Noise Ratio (SNR) and QAM side (number of quantization levels per QAM axis) for the Deep Signal Network~\cite{tu2020complex} architecture. Outlined region indicates combinations with $\le5\%$ drop in accuracy, where the digitally trained accuracy was 68\%.
    }
    \label{fig:accuracy_degradation_radioml}
\end{figure}

Decreasing constellation size (measured by QAM side) and hardware SNR will both degrade inference accuracy. Comparing to a digitally-run network with infinite SNR and essentially infinite constellation size, we plot the accuracy degradation under different combinations of constellation size and SNR in Figure \ref{fig:accuracy_degradation_radioml}.
The outlined area in the plot, highlighting the combinations that result in less than $5\%$ loss in accuracy, is roughly the regime of QAM side $\ge\!\!32$ levels and $\ge\!\!20$~db of SNR. These results suggest that implementing our I/Q photoelectric multiplier using only conventional, off-the-shelf QAM modulators (that typically have over 20 dB of SNR) will not significantly sacrifice accuracy. Using \mbox{256-QAM} modulators (16 levels per side of modulation) and a moderate 30 dB of SNR results in an accuracy drop of only~7.3\%. While accuracy degradation from quantization and noise is highly model-dependent~\cite{nagel2021white}, these findings indicate the broader feasibility of deploying off-the-shelf deep complex-valued neural networks using commodity telecommunications hardware.

\section{QAMNet: Model-Hardware Trade-Offs for QAM-Based Optical Neural Networks}\label{sec:tradeoffs_and_equivalences}
\begin{table*}
    \centering
    \begin{tabular}{m{3cm}cccc} \toprule
         & QAMNet (this work) & Level Equivalent 1D & Hardware Equivalent 1D & \makecell[c]{Energy Equivalent 1D \\ (\# values $\times$ energy per value)} \\\midrule
         
        \# Total Levels & $N$ & $N$ & $\sqrt N$ & $\left\lceil \sqrt{2(\sqrt{N}-1)^2}\right\rceil + 1$ \\
        
        \# Bits per Value & $\log_2(N)/2$ & $\log_2(N)$ & $\log_2(N)/2$ & $\log_2\left(\left\lceil \sqrt{2(\sqrt{N}-1)^2}\right\rceil + 1\right)$ \\
        
        \# Weight Values & $2(w^2h+h)+2(hc+c)$ & $w^2h+h+hc+c$ & $w^2h+h+hc+c$ & $w^2h+h+hc+c$ \\
        
        Energy Per Value ($\Delta^2$) & $\left(\left(\sqrt N - 1\right)/2\right)^2$ & $\left((N-1)/2\right)^2$ & $\left(\left(\sqrt N - 1\right)/2\right)^2$ & $2\left(\left(\sqrt N - 1\right)/2\right)^2$ \\

        \smash{\raisebox{9mm}{Example with $N=16$}} & 
        \includegraphics[width=0.15\linewidth]{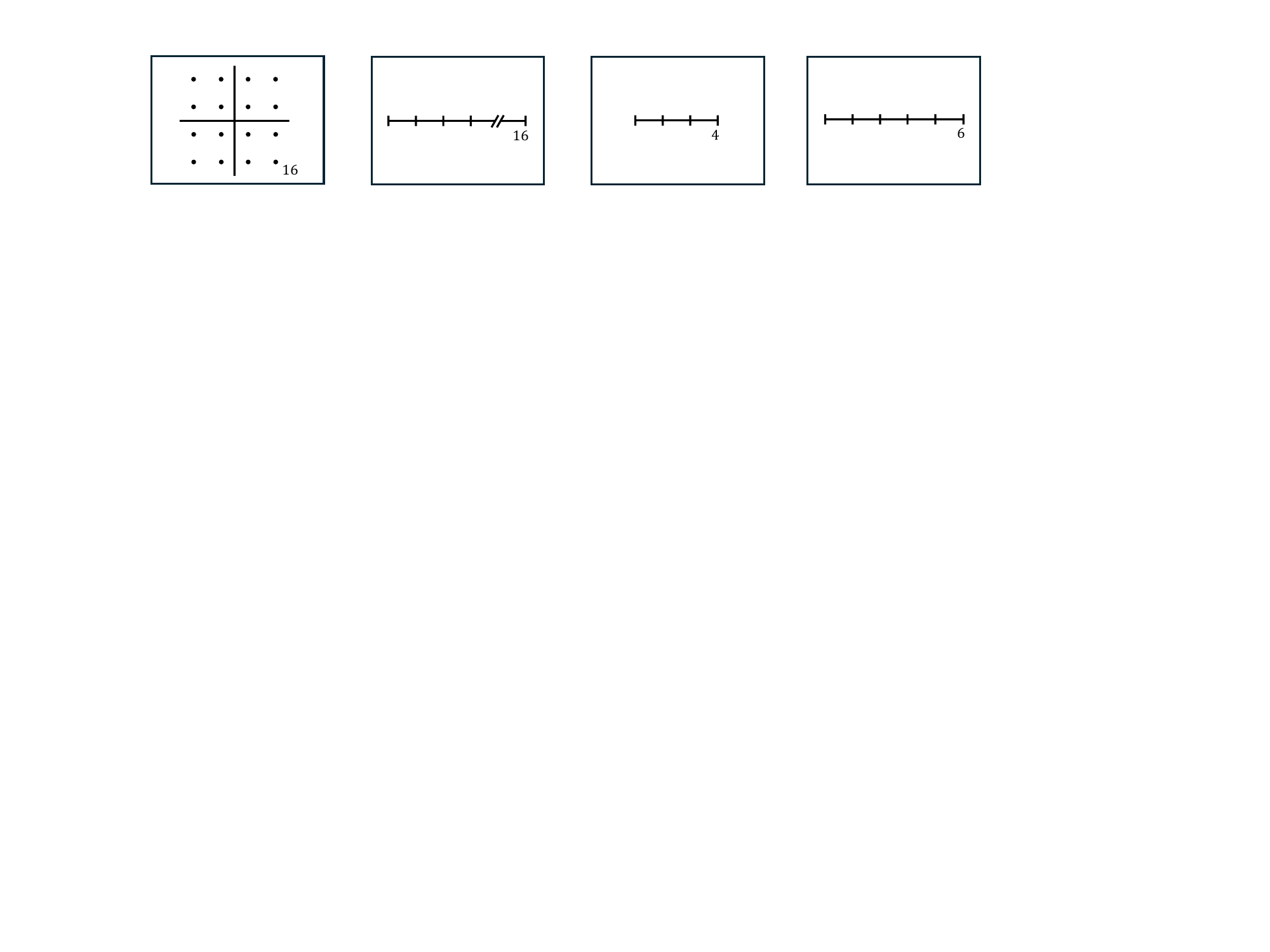} & 
        \includegraphics[width=0.15\linewidth]{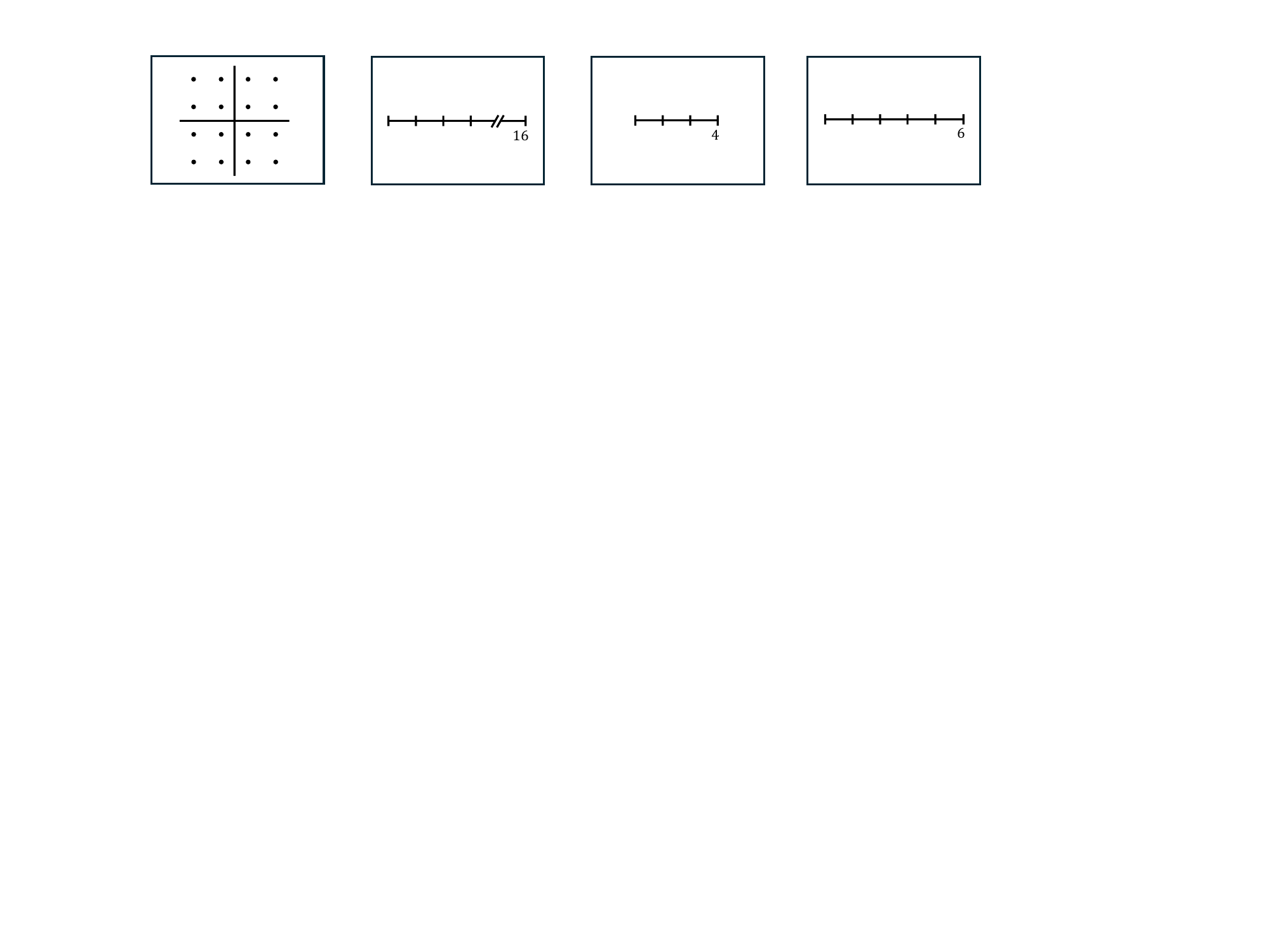} &
        \includegraphics[width=0.15\linewidth]{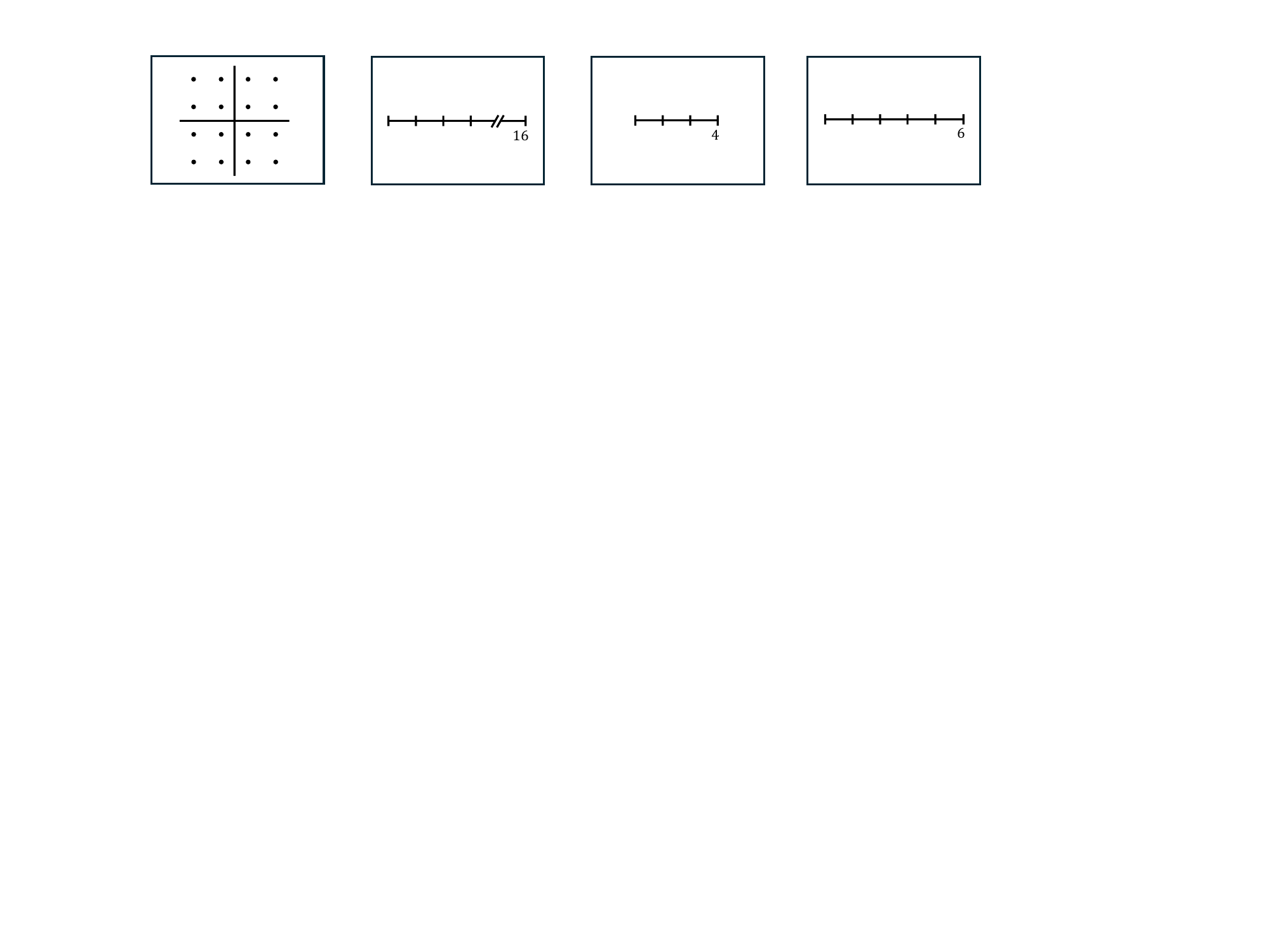} &
        \includegraphics[width=0.15\linewidth]{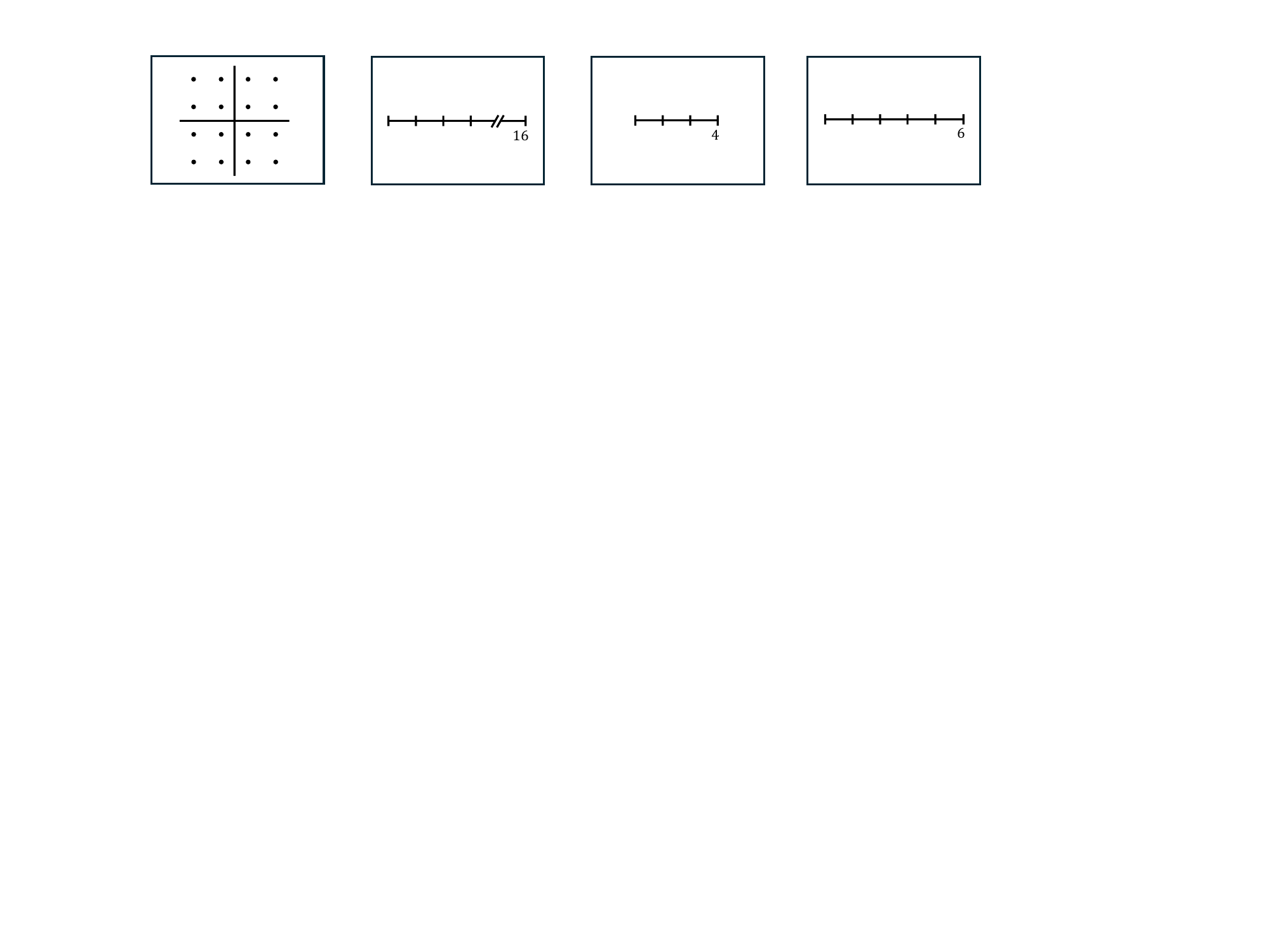} \\ 

        \bottomrule
        
    \end{tabular}
    \caption{Model and hardware tradeoffs of a 1D (real-valued) neural network against a complex-valued neural network (QAMNet). All networks take a $w^2$ size input, have one hidden layer of $h$ neurons, and produces a classification output of $c$ classes. Each equivalent 1D network holds a different metric constant equal to the QAMNet. Energy Per Value is in units of $\Delta^2$, where $\Delta$ is the minimum distinguishable step size between two amplitude values.}
    \label{tab:equivalences}
\end{table*}

In Section \ref{sec:inference_with_telecom_qam_modulators}, we demonstrated the feasibility of implementing pre-trained complex-valued neural networks (CVNNs) on QAM hardware for complex-valued datasets. In this section, we introduce the \textit{QAMNet} architecture of complex-valued neural networks implemented with I/Q photoelectric multiplication, and compare it against real-valued neural networks implemented with amplitude photoelectric multipliers (\textit{1D~ONNs}). 
QAMNet uses standard QAM modulators with finite precision in the I/Q photoelectric multiplication scheme, while 1D ONNs use a single axis of amplitude modulation. We use real-valued datasets for this comparison.

\subsection{QAMNet}
Because the MNIST family of image classification tasks contain real-valued inputs, QAMNet uses a trainable embedding layer, inspired by word2vec \cite{mikolov2013efficient}, to map each real-valued input pixel value to a complex number. These complex values are then processed with a complex-valued neural network implemented with I/Q photoelectric multipliers as discussed in Section \ref{sec:photoelectric_multiplication_scheme}.
The mapping from $\mathbb R^d\to\mathbb C^d$ is expressed as repeatedly applying ($d$ times) the map $\mathbb R\to\mathbb C$, which is structurally equivalent to applying the map $\mathbb R\to\mathbb R^2$. In the case of MNIST-style images as shown in Figure \ref{fig:main_figure}.b., this map corresponds to an embedding from a vocabulary of length 256 (the total number of possible pixel values) to a vector space of dimension 2. Embeddings like this can be readily implemented with a lookup table that requires no MAC operations, making this a feasible approach for edge devices. During training, the embedding layer is differentiable and is trained alongside the rest of the network using backprop.

\subsection{Evaluation Principles}
One particularly promising use case of optical neural networks is in low-power edge device intelligence, using schemes such as Netcast \cite{hamerly2022netcast}. Edge devices are characterized by tight constraints on cost, energy, and memory, and often, the desire for low latency and high throughput neural network inference. In edge devices operating under low memory constraints, any data on the edge device must be minimized, including both the input data size and the layer activations that are stored during inference.

In the context of edge devices, we study MLPs with low neuron counts and two layers. The latency of inference scales linearly with the number of neurons, motivating the study of neurons with low neuron counts. The amount of storage required on the edge device grows with the size of the intermediate layer activations that must be stored between the computation of one layer and the next. 
With the same motivations, we benchmark on the MNIST~\cite{lecun1998gradient} dataset downsampled to $7\times7$ pixels. We include benchmarks on the variants of Fashion~MNIST~\cite{xiao2017fashion} and KMNIST~\cite{clanuwat2018deep} in Appendix~\ref{appendix:other_datasets}. 

\subsection{Energy Consumption of Activations}\label{sec:Energy Consumption of Activations}
In Netcast\cite{hamerly2022netcast}, a client performing DNN inference receives weights streamed from a centralized weight server, and uses the transmitted weights directly in photoelectric multiplication. As a result, a client need only modulate the DNN's input and each layer's input activations, dramatically reducing its onboard energy and memory requirements. Consider a multi-layer perceptron (MLP) with $L$ layers, with the number of neurons at layer defined as $h$, where $h_0$ is the input dimension, $h_i$ for $1\leq i\leq L$ is the number of neurons at the $i$'th hidden layer, and $h_{L+1}$ is the number of classes in the output layer. For a 1D ONN and QAMNet with $N$ total levels, the total client energy consumption is the number of values to be modulated multiplied by the energy per modulation:
\begin{align}
    \text{1D ONN Energy: }&\left(\sum_i h_i\right)\times\left(\frac{N-1}{2}\right)^2\\
    \text{QAMNet Energy: }&2\times\left(\sum_i h_i\right)\times\left(\frac{\sqrt N - 1}{2}\right)^2
\end{align}

\subsection{Terminology for Comparison}
When picking the parameters for a 1D ONN to compare against QAMNet, there are multiple design decisions that can be explored, depending on what one chooses to keep constant. We investigate the three comparison classes in Table~\ref{tab:equivalences}.

We use the term \textit{value} to refer to a single real number used in a network's parameterization. In QAMNet, we count each complex-valued weight as being composed of two values. In a 1D ONN, a single real-valued weight corresponds to a single value. Each weight is modulated with a number of \textit{total levels}. In QAMNet, the total number of levels $N_\text{total}=N_\text{side}^2$ is the number of constellation points, where $N_\text{side}$ is the number of levels realizable by a single modulator. In 1D ONNs, the total number of levels is simply the number of levels realizable by the single modulator. Each modulator can realize a certain number of levels, whose equivalent bit precision is denoted as \textit{\# Bits per value}. The maximum power required for a single amplitude modulator to represent a single value, which is the square of the maximum amplitude. Because the minimum spacing of amplitudes is lower bounded by the noise $\Delta$ around each symbol, for a number of levels $L$ of a single value, the \textit{energy per value} is computed as $((L-1)/2)^2\Delta^2$.

\subsection{Results}\label{sec:Results}
\begin{figure*}
    \centering
    \includegraphics[width=1\linewidth]{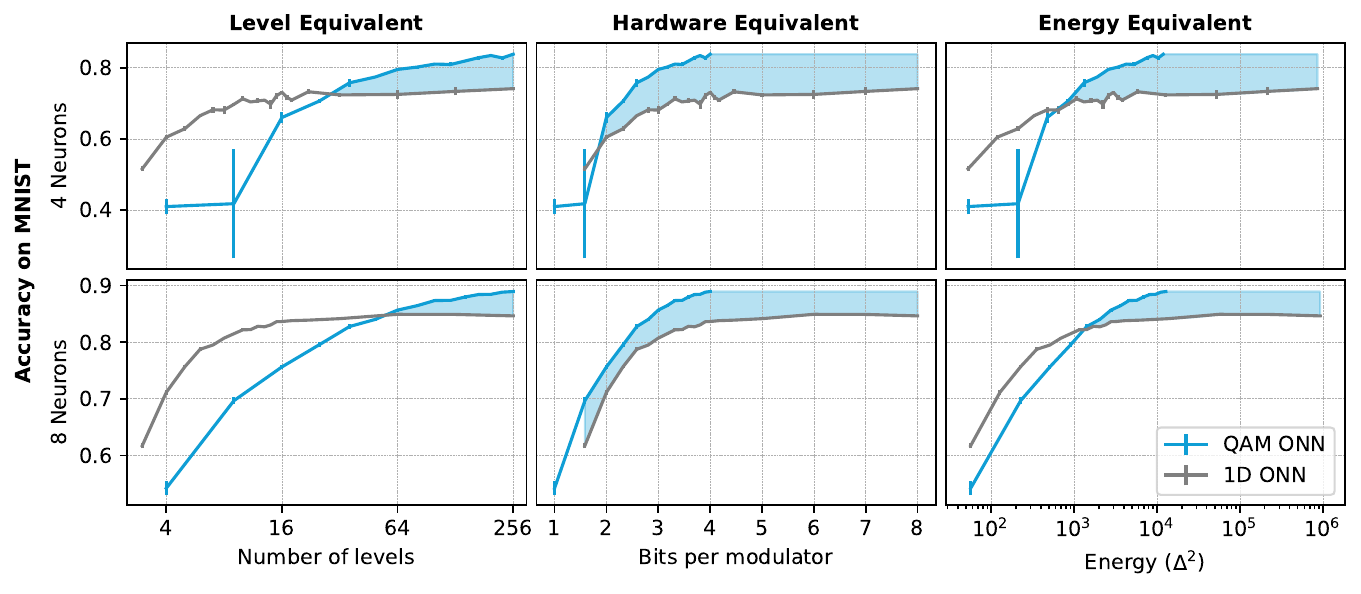}
    \caption{Accuracy comparisons between QAMNet and level equivalent, hardware equivalent, and energy equivalent 1D ONNs on the MNIST dataset. Equivalence definitions found in Table \ref{tab:equivalences}. The shaded blue region highlights the regimes of advantage of QAMNet over the equivalent 1D ONN, achieving up to $9.7\%$ greater accuracy.}
    \label{fig:results_plot}
\end{figure*}

\subsubsection{Level Equivalent Comparison}\label{sec:Level Equivalent}
% As the total number of levels increase, QAMNet achieves higher accuracy than 1D ONN counterparts while using quadratically less energy for the same number of levels. 
This equivalence perspective highlights the relationship between quantization levels in 1D ONNs compared to quantization levels in QAMNet.
% This equivalence perspective highlights the relationship between quantization levels in one dimension (amplitude) compared to quantization levels in two dimensions (I/Q modulation via QAM). 
 %For a QAM ONN that uses a $N$-QAM constellation with $N$ unique points, the level-equivalent 1D ONN uses a single modulator that has $N$ levels. As a result, the QAM ONN uses quadratically less energy. To realize $N$ total levels with two modulators, the energy is proportional to $\left(2\sqrt{N}\right)\Delta^2$, while the energy required to modulate $N$ unique levels with one modulator is proportional to $N\Delta^2$.
At extremely low numbers of levels, the 1D ONN achieves higher accuracy than QAMNet, likely because gradients computed during backprop become significantly inaccurate with a low number of levels per value. 
%$N$ total levels in QAM correspond to quantizing each complex-valued weight component to $\sqrt N$ levels, introducing significantly more quantization error per value than the $N$ quantization levels of 1D ONNs. Lower numbers of levels correspond to higher quantization error and increased error in gradient estimates, causing backprop to be less robust as the training problem approaches integer optimization. 
However, as the number of levels increase and accuracy is not limited by quantization effects on training, QAMNet achieves higher accuracies that 1D ONNs are never able to match. The crossover point seems to move to the right as the network size increases, suggesting that the advantage of QAMNet is most significant in low power and low precision hardware environments.
% \textbf{The level equivalence perspective shows that for a fixed network architecture and size, QAM ONNs yield higher accuracies while using quadratically less power, above a certain threshold of bit precision.}

\subsubsection{Hardware Equivalent Comparison}
% QAMNet is an optimal strategy when hardware precision is limited. They consistently achieve higher accuracy than even 1D ONNs that have far greater modulator precision. 
In low-power and low-cost devices, SWAP, cost and supply chain limitations often restrict the maximum precision of available modulators. In these scenarios, QAMNet consistently matches or surpasses the performance of traditional 1D ONNs. 
For a fixed bit precision per amplitude modulator, QAMNet can achieve superior performance by using two modulators together for I/Q photoelectric multiplication, compared to a 1D ONN that uses one amplitude modulator. 
%Even though the QAM ONN has twice as many weight values, the QAM ONN's scaling of total levels to energy is quadratically better. 
Linearly increasing the number of levels of a modulator results in a linear increase in total levels in 1D ONNs, but a quadratic increase in total levels in QAMNet. As a result, QAMNet consistently achieves higher accuracy than even 1D ONNs that have far greater modulator precision. 

\subsubsection{Energy Equivalence Comparison}
% \textbf{QAM ONNs achieve equal or higher accuracies than 1D ONNs for the same energy, above a certain energy threshold.}
% QAMNet is an optimal strategy to maximize accuracy, above a certain problem- and architecture-dependent energy threshold.
In this comparison, the total energy of the 1D ONN matches that of QAMNet by increasing the number of levels in the 1D ONN to the next largest integer value (see formula in ``Energy Equivalent 1D" in Table~\ref{tab:equivalences}).
The total energy is computed as the number of activation values multiplied by the power required to realize each value, as described in Section \ref{sec:Energy Consumption of Activations}. 
At extremely low energy budgets, QAMNet performs slightly worse than 1D ONNs due to training instability and sub-optimal convergence with low precision quantization, but above moderate energy budgets, QAMNet demonstrates a clear accuracy advantage over 1D ONNs. For the same energy per value of $\left(\left(\sqrt{N}-1\right)/2\right)^2\Delta^2$, QAMNet can represent $N$ total levels, while the hardware equivalent 1D ONN can only represent $\sqrt N$ total levels. As a result, for the same energy utilization, QAMNet achieves overall higher accuracies than 1D ONNs. 
%\textbf{The energy equivalent perspective shows that QAM ONNs achieve equal or higher accuracies above a certain architecture-dependent energy threshold.}

\section{Conclusion}
We have presented a photoelectric multiplication scheme that uses I/Q modulated signals to perform efficient complex-valued inner products. This scheme can be readily implemented with telecommunications-grade QAM modulators to accurately accelerate complex-value neural network inference, indicating an opportunity for efficient edge device DNN inference using existing telecommunications equipment.

We then benchmarked an end-to-end information processing system that maps real-valued inputs to complex values, which are then processed with a complex-valued neural network implemented on QAM hardware. The three perspectives of level equivalent, hardware equivalent, and energy equivalent demonstrate different regimes of advantage of QAMNet over 1D counterparts. As discussed in Section~\ref{sec:Results}:

\begin{itemize}
    \item The \textbf{Level Equivalent} comparison reveals that QAMNet is significantly more energy-efficient than their 1D counterparts: for an equivalent number of total levels, QAMNet uses quadratically less energy. This efficiency is attributed to the ability of QAMNet to utilize two modulators in parallel, achieving the same number of unique levels with reduced power consumption. Consequently, QAMNet achieves higher accuracy with less energy, making them a preferable choice in energy-constrained scenarios.

    \item The \textbf{Hardware Equivalent} comparison demonstrates that when hardware bit precision is limited, QAMNet consistently demonstrates superior performance by leveraging two modulators for I/Q modulation. QAMNet outperforms 1D ONNs in terms of accuracy, due to the quadratic increase in total levels achievable by QAMNet compared to the linear increase in 1D ONNs. Thus, QAMNet is an optimal strategy for low-power and low-cost devices where hardware precision is a constraint.

    \item The \textbf{Energy Equivalent} comparison further illustrates the efficiency of QAMNet, showing that as the total energy allotment increases, QAMNet achieves higher accuracies than 1D ONNs. Using the same amount of energy for total levels is more useful in QAM than in 1D ONNs. 
    
\end{itemize}

% Our analysis demonstrates that QAM ONNs offer substantial benefits over 1D ONNs across various metrics, including energy efficiency, performance with low-precision hardware, and accuracy. By utilizing both axes of I/Q modulation, QAM ONNs can achieve superior performance with less power. These findings suggest that QAM ONNs are a highly efficient and effective approach for neural network inference under limited energy and hardware precision constraints.

In summary, our analysis demonstrates that QAMNet offers substantial benefits over 1D ONNs across various metrics. QAM-based ONNs like QAMNet can achieve higher accuracy at lower energies for a given bit precision, attain higher overall accuracies above moderate energy budgets, and are an optimal choice when hardware bit precision is limited. As DNN intelligence becomes increasingly prominent on edge devices and in power-constrained environments, QAM-based ONNs offer a compelling balance of efficiency and performance.

% As new computing paradigms emerge from novel computing substrates like photonics, findings like this one suggest the opportunities for energy and performance benefits that may arise from co-designing hardware and software together. 

\section{Methods}

% \subsection{Post-Training Quantization}
In Section \ref{sec:inference_with_telecom_qam_modulators}, post-training quantization of the CVNN is performed by finding scaling and zero-point factors for each inner product, which corresponds to independent factors for each neuron. Quantization is simulated by scaling and shifting to the range $[-1, 1]$, applying the uniform quantizer to those values, and then computing the unscaled output. 

% \subsection{Quantization-Aware Training} 
The QAMNet and 1D ONN models are trained using quantization-aware training (QAT) as described in \cite{jacob2018quantization}, where weights and inputs are quantized to the QAM constellation during the forward pass, while full precision weights and inputs are used to compute gradients in the backward pass based on the error of the quantized weights and inputs. For small networks, we find that this yields better accuracies than post-training quantization. 

During backprop, gradients of the quantization function are computed with the straight-through estimator~\cite{bengio2013estimating}. In the forward pass, weights $W$ and inputs $x$ are quantized with a function $Q$, such each layer's post-activation output is computed as  
\begin{equation}
    y=f(Q(W)Q(x^*))
\end{equation}
where $f$ is the desired nonlinear activation. %In the backwards pass, the gradient of the non-differentiable quantization function is defined as \mbox{$\frac{\partial}{\partial x}(Q(x))\approx 1$}. 
In Section \ref{sec:tradeoffs_and_equivalences}, quantization-aware training is simulated as a uniform quantizer with a fixed dynamic range of $[-1, 1]$. Specifically, the derivative of the quantization function is set to be
\begin{equation}
    \frac{\partial}{\partial x} Q(x)\approx\begin{cases}
        1&\text{if}~-1\le x\le 1\\
        0&\text{otherwise}
    \end{cases}
\end{equation}

The smallest network sizes and lowest precision quantization (4 hidden neurons and 4-16 levels) experience some fluctuations in accuracy due to training instability arising from the effect of weight oscillations, a well-known phenomena when using very low bit precision quantization-aware training~\cite{nagel2022overcoming}. Nevertheless, these do not affect the overall results, and their effect diminishes at higher levels of quantization.

% \subsection{Simulation}
Simulations were developed in Tensorflow, and were run on the Massachusetts Green Energy High Performance Computing Center (MGHPCC)'s Engaging cluster. Weights and Biases was used to collect and organize results.

\section*{Acknowledgment}

We would like to thank Ryan Hamerly and Saumil Bandyopadhyay for their insightful comments and discussion. K. S. acknowledges the support of the Israeli Council for Higher Education and the Zuckerman STEM Leadership Program.

\bibliographystyle{IEEEtran}
\bibliography{refs}

\begin{appendices}
\section{I/Q-Based Complex-Valued Inner Products}\label{appendix:iq_inner_products}

In considering the workings of the I/Q inner product engine, the top and bottom mixer beam paths of Figure \ref{fig:main_figure} behave similarly. We will omit vector arrow superscripts and denote their real and imaginary parts with superscripts $^{(r)}$ and $^{(i)}$ for notational simplicity.

The two I/Q modulated passband signals entering the homodyne demodulator are \mbox{$s_1(t)=(w_j^{(r)}+iw_j^{(i)})e^{i\omega t}$} and \mbox{$s_2(t)=(x_j^{(r)}+ix_j^{(i)})e^{i\omega t}$}, where $\omega$ is the carrier frequency. In the upper beam path, the mixer's effective transition matrix combines a beamsplitter matrix and a $90^\circ$ phase shift on one input port. For the $j$'th element of the inner product $w\cdot x^*$, the signals entering the photodetectors are
\begin{equation}
\begin{split}
\frac{1}{\sqrt{2}}\begin{bmatrix}1&1\\1&-1\end{bmatrix}\begin{bmatrix}i&0\\0&1\end{bmatrix}\begin{bmatrix}
(w_j^{(r)}+iw_j^{(i)})e^{i\omega t}\\
(x_j^{(r)}+ix_j^{(i)})e^{i\omega t}\\
\end{bmatrix}\\
=\frac{1}{\sqrt 2}\begin{bmatrix}
(x_j^{(r)}-w_j^{(i)}+i(w_j^{(r)}+x_j^{(i)}))e^{i \omega t}\\
(-x_j^{(r)}-w_j^{(i)}+i(w_j^{(r)}-x_j^{(i)}))e^{i \omega t}\\
\end{bmatrix}
\end{split}
\end{equation}

From each output port of the homodyne mixer, the photocurrents for the $j$'th element of the inner product $w\cdot x^*$ are the squared output intensities.
\begin{equation}\label{eq:photocurrent_single_element}
\begin{split}
    I_j^+&=\frac{1}{2}\left\|x_j^{(r)}-w_j^{(i)}+i(w_j^{(r)}+x_j^{(i)})\right\|^2\\
    I_j^-&=\frac{1}{2}\left\|-x_j^{(r)}-w_j^{(i)}+i(w_j^{(r)}-x_j^{(i)})\right\|^2
\end{split}
\end{equation}

Summing the difference of these intensities over the elements of $j$ yields the desired inner product:
\begin{equation}
\begin{split}
    \sum_j I_j^+-I_j^-&=\sum_j 2\left(w_j^{(r)}x_j^{(i)}-w_j^{(i)}x_j^{(r)}\right)\\
    &=2\text{Im}(w\cdot x^*)
\end{split}
\end{equation}

In the bottom beam path, the same analysis can be applied, omitting the $90^\circ$ phase shift, yielding the result $2\text{Re}(w\cdot x^*)$.

\section{Doubly Fast Real-Valued Dot Products} \label{sec:doubly_fast_real}
The inner product of two real-valued length-$n$ vectors \mbox{$\vec a=[a_1, a_2, \dots,a_n]$} and \mbox{$\vec b=[b_1, b_2,\dots,b_n]$} can be expressed as the sum of $n/2$ inner products of length-$2$ vectors: 
\begin{align}\label{eq:decompose_dot_product}
\vec a \cdot \vec b&=\sum_{i=1}^{n/2}\left(a_{2i}b_{2i}+a_{2i+1}b_{2i+1}\right)\\
&=\sum_{i=1}^{n/2}\text{Re}\left((a_{2i}+ia_{2i+1})(b_{2i}+ib_{2i+1})^*\right)
\end{align}
The inner product of two length-$2$ vectors $[a, b]\cdot[c, d]$ can be computed as the the real component of the complex-valued multiplication, $\text{Re}((a+ib)(c+id)^*)$. This computation happens in a single time step at one mixer's output of the I/Q photoelectric multiplier. Therefore, by ``rolling" length-$n$ real-valued vectors into $n/2$ length-2 vectors by using every other value as the complex component of its preceding real value, the inner product of two length-$n$ vectors can be computed in $n/2$ time steps using the I/Q photoelectric multiplier.

\section{Comparison to Alternative Designs}\label{appendix:alternative_designs}
\begin{table*}%[h]
\begin{center}
\resizebox{\textwidth}{!}{
\begin{tabular}{>{\centering\arraybackslash}m{0.0\linewidth} >{\centering\arraybackslash}m{0.33\linewidth} >{\centering\arraybackslash}m{0.33\linewidth} >{\centering\arraybackslash}m{0.33\linewidth}} \toprule
 & \textbf{I/Q Photoelectric Multiplication} & \textbf{4 Inner Product Engines} & \textbf{2 Mixers} \\
 & Complex-Valued & Real-Valued & Real-Valued \\ \midrule
& \includegraphics[width=\linewidth]{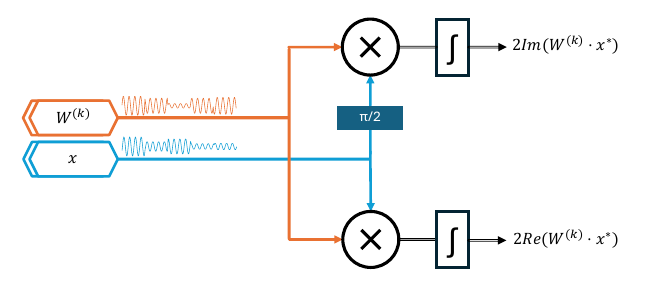} & 
\includegraphics[width=\linewidth]{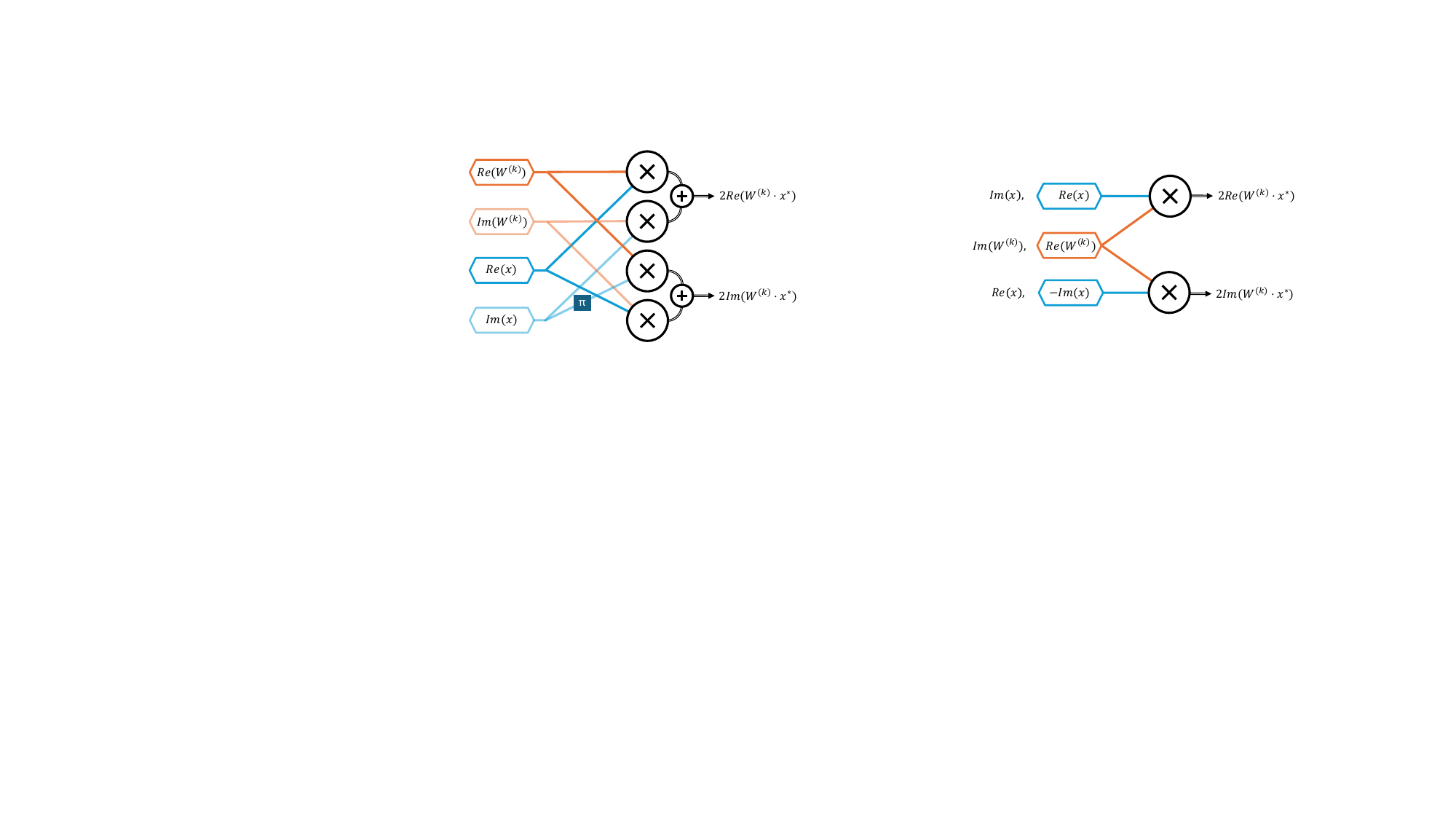} & 
\includegraphics[width=\linewidth]{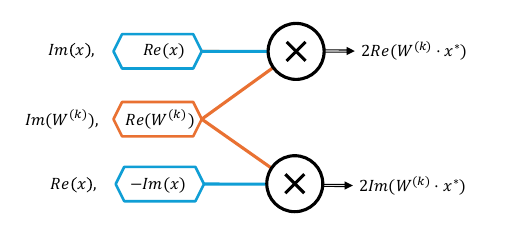} \\ \midrule
\textbf{Modulators} & 4 (2 per I/Q modulator) & 4 & 3 \\ 
\textbf{Mixers} & 2 & 4 & 2 \\
\textbf{Timesteps} & 1 & 1 & 2 \\
\bottomrule
\end{tabular}
}
\end{center}
\caption{Comparison of I/Q Photoelectric Multiplication with alternative amplitude-only schemes for complex-valued inner products. The number of timesteps is shown for computing a single complex-valued multiplication of $W^{(k)}x^*$.}
\label{tab:other_designs}
\end{table*}

Table \ref{tab:other_designs} compares the I/Q photoelectric multiplier with alternative designs for complex-valued inner product engines that use amplitude-only modulation. The naive approach may be to use 4 distinct mixers representing the partial sums $ac$, $bd$, $-ad$, and $bc$, with each operand modulated once and fanned out into two mixers. Note that in order to achieve the negative sign on the $-ad$ term, a $\pi$ phase shift is applied on the beam path of $\text{Im}(x)$. The partial sum terms are then added in analog electronics, yielding the real and imaginary components when summed together. Alternatively, a complex-valued multiplier can be built with two mixers and three modulators, but the design requires two time steps to accumulate the real and imaginary components. Our proposed I/Q photoelectric multiplier uses two mixers and four modulators for one time step, resulting in the most efficient implementation.

\section{Other Datasets}\label{appendix:other_datasets}
\begin{figure}
    \centering
    \includegraphics[width=1\linewidth]{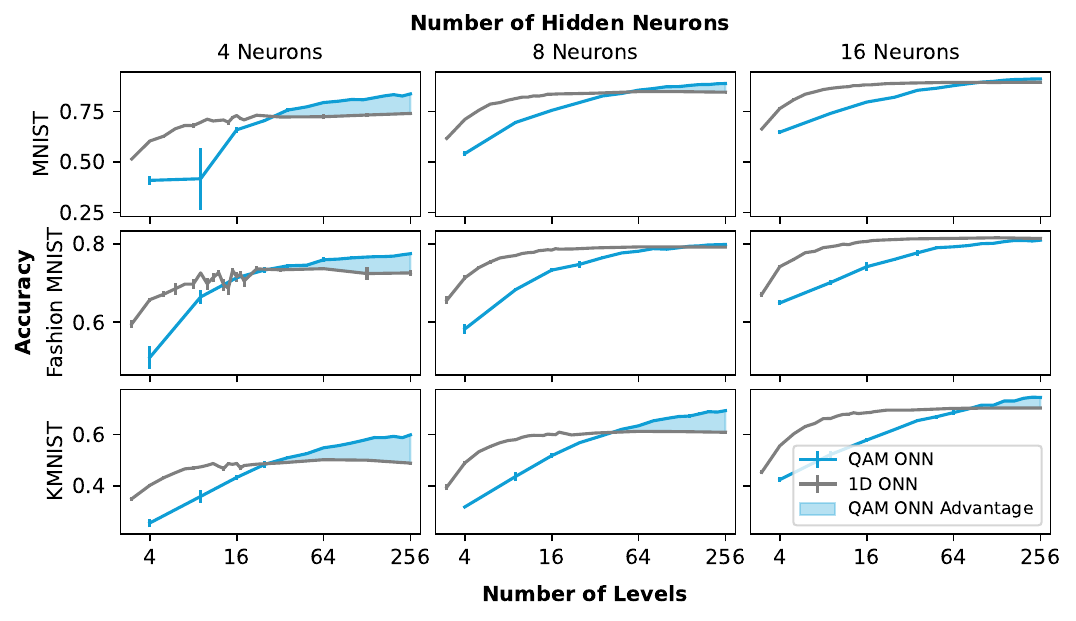}
    \caption{QAM ONNs vs. Level Equivalent 1D ONNs, for different numbers of total levels (QAM constellation points). For an equivalent number of total levels, the QAM ONN uses quadratically less energy than the 1D ONN.}
    \label{fig:level_equivalence}
\end{figure}

\begin{figure}
    \centering
    \includegraphics[width=1\linewidth]{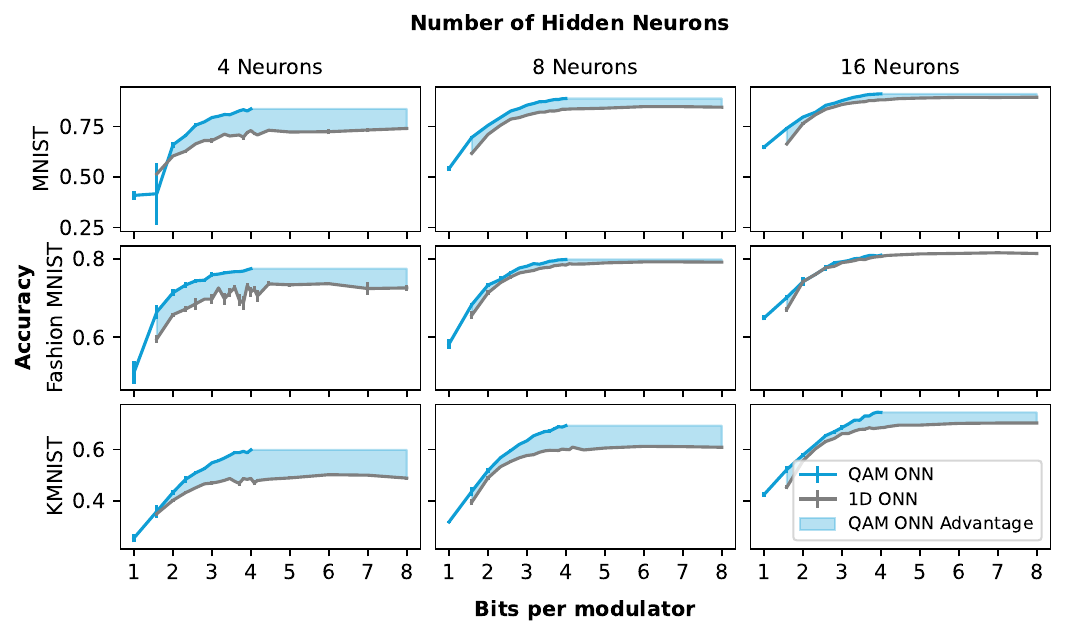}
    \caption{QAM ONNs vs. Hardware Equivalent 1D ONNs that have the same bit precision per modulator. Given amplitude modulators of limited bit precision, the QAM ONN enables higher accuracy using two of them in parallel for I/Q modulation.}
    \label{fig:hardware_equivalence}
\end{figure}

\begin{figure}
    \centering
    \includegraphics[width=1\linewidth]{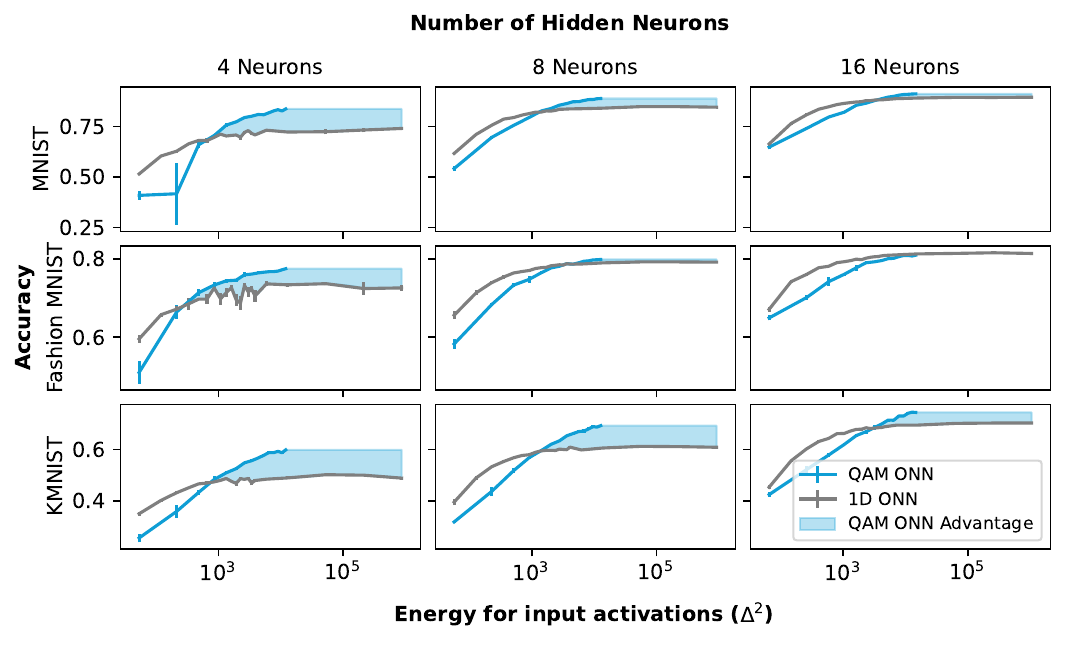}
    \caption{QAM ONNs vs. Energy Equivalent 1D ONNs that have the same total inference energy, accounting for input activation energy. At moderate energy budgets and above, QAM ONNs achieve higher accuracy.}
    \label{fig:energy_equivalence}
\end{figure}

In addition to MNIST, we include benchmarks on the similar datasets Fashion~MNIST~\cite{xiao2017fashion} and KMNIST~\cite{clanuwat2018deep} for 2-layer networks with 4, 8, and 16 neurons in the hidden layer. Some training instability arises from weight oscillations \cite{nagel2022overcoming} at few neurons and low levels of quantization. We present full results for the level equivalence (Figure \ref{fig:level_equivalence}), hardware equivalence (Figure \ref{fig:hardware_equivalence}), and energy equivalence (Figure \ref{fig:energy_equivalence}) perspectives.

\end{appendices}
\end{document}